\documentclass[10pt, switch, twocolumn]{article}

\usepackage{arxiv}

\usepackage{amsmath}
\usepackage{amsfonts}
\usepackage{amssymb}
\usepackage{lmodern}
\usepackage{xcolor,soul}
\usepackage{graphicx}
\usepackage{physics}
\usepackage[utf8]{inputenc}
\usepackage{subfig}
\usepackage{siunitx}
\usepackage[shortlabels]{enumitem}
\usepackage{float}
\usepackage{array}
\usepackage{booktabs}
\usepackage{fancyhdr}
\usepackage{caption}

\DeclareSIUnit \haunit{Ha}

\usepackage[hidelinks]{hyperref}
\usepackage[numbers,square]{natbib}
\hypersetup{
    colorlinks = true,
    urlcolor = {blue},
    citecolor = {.},
    linkcolor = {.},
    anchorcolor = {.},
    filecolor = {.},
    menucolor = {.},
    runcolor = {.},
    pdftitle={}, 
    pdfsubject={}, 
    pdfauthor={}, 
    pdfkeywords={}, 
}
\title{
    Chemically Aware Unitary Coupled Cluster with 
    \textit{ab initio} Calculations on System Model H1: 
    A Refrigerant Chemicals' Application
    }

\author{
    \textbf{I. T. Khan, M. Tudorovskaya, J. J. M. Kirsopp, D. Mu\~{n}oz Ramo} \\
    Quantinuum, Terrington House, 13–15 Hills Road, Cambridge CB2 1NL, United Kingdom \\
    \newline \\
    \textbf{P. Warrier, D. K. Papanastasiou, R. Singh} \\
    Honeywell Advanced Materials, 20 Peabody St, Buffalo, NY 14210, United States \\
}

\begin{document}
\urlstyle{rm} 

\pagenumbering{roman}    
\pagenumbering{roman}

\pagenumbering{arabic}

\twocolumn[ 
  \begin{@twocolumnfalse} 
    \maketitle

    \date{\today}

    \begin{abstract}
        Circuit depth reduction is of critical importance for quantum chemistry simulations on current and near 
        term quantum computers. This issue is tackled by introducing a chemically aware strategy for 
        the Unitary Coupled Cluster ansatz. The objective is to use the chemical description of a system to aid 
        in the synthesis of a quantum circuit. We combine this approach with two flavours of Symmetry 
        Verification for the reduction of experimental noise. These method enable the use of System Model 
        H1 for a 6-qubit Quantum Subspace Expansion calculation. We present $(i)$ calculations to obtain 
        methane's optical spectra; $(ii)$ an atmospheric gas reaction simulation involving 
        {[CH$_3^{\cdot}$ --- H --- OH]}$^{\ddagger}$. Using our chemically aware unitary coupled cluster state-preparation strategy in tandem 
        with state of the art symmetry verification methods, we improve device yield for CH$_4$ at 6-qubits. 
        This is demonstrated by a 90\% improvement in two-qubit gate count and reduction in relative error to 0.2\% for 
        electronic energy calculated on System Model H1.
    \end{abstract}
    \vspace{0.4cm}
    \vspace{0.4cm}    
    \end{@twocolumnfalse}
]

\label{sec::introduction}
The simulation of molecules and materials using quantum chemistry methods is a well established field, 
with applications in many scientific and industrial areas of interest \cite{RevModPhys.71.1253}. 
However, there is awareness about the shortcomings of performing these simulations on classical machines \cite{Cao_2019}. 
The main workhorse of classical simulations, Density Functional Theory, fails to capture the qualitative 
behaviour of chemicals with strong correlation \cite{dft_perspective}. One naturally turns to wave function 
methods, such as Hartree-Fock (HF), Coupled Cluster (CC) theory and Configuration Interaction (CI) methods. 
However, the steep memory requirements of these techniques have limited their practicality in the study of 
complex chemicals. This issue also plagues classical excited state heuristics that provide CI- or 
CCSD-like accuracy. For instance, Equation of Motion Coupled Cluster (EOM-CCSD) scales as 
$O({N^6})$, where $N$ is the number of spin orbitals \cite{eom_ccsd}.

Quantum computers provide an alternative path to enable the use of wave function methods on problems of 
practical interest. In particular, two state of the art algorithms have been proposed, the Variational 
Quantum Eigensolver (VQE) and Quantum Subspace Expansion (QSE) algorithms, as the earliest candidates 
for scalable alternatives to classical wave function methods \cite{Peruzzo_2014, Kandala_2017, McClean_2017}. 
VQE approximates the ground-state of a Hamiltonian variationally, and QSE estimates excited states 
non-variationally. Despite the rapid development and deployment of quantum resources, progress has 
been limited due to quantum hardware being unable to meet performance requirements of these algorithms. 
Nevertheless, the exploration of quantum computing applications for chemistry is necessary for the 
progression of computational chemistry, and also serve as important benchmarks for today's hardware 
and its relevance to the chemicals industry. 

Quantum algorithms for chemistry require a large amount of quantum resources for the state-preparation 
component of these simulations.  Broadly, two families of state-preparation methods have been introduced 
in literature: Unitary Coupled Cluster (UCC), and hardware-efficient ans{\"a}tze. With UCC, the number of 
variational circuit parameters and the 2-qubit gate depth scale as $O({N^4})$, leading to circuits beyond the 
capability of today's machines \cite{McClean_2016}. Hardware-efficient methods require less coherent resources, 
but suffer from the ``barren plateau'' problem \cite{McClean2018}. Initial effort has been focused on 
improving the 2-qubit depth scaling for the UCC ansatz. Notably, adaptive methods 
\cite{https://doi.org/10.48550/arxiv.1809.03827}, circuit recompilation 
\cite{Jones2022robustquantum}, and unique UCC circuit synthesis approaches \cite{Cowtan_2020}. 
Adaptive methods are unattractive. For example, ADAPT-VQE requires additional measurements on top of regular 
VQE \cite{Grimsley_2019}. This encourages us to explore a unique UCC circuit synthesis approach that improved 
the 2-qubit gate count by describing Fermionic UCC spatial orbital to spatial orbital excitations as hard-core 
Bosonic operators \cite{nam2019groundstate, Kottmann_2022, Elfving_2021}. In tandem, there has been additional 
work on discarding UCC excitations by using symmetry filtering as a priori 
\cite{https://doi.org/10.48550/arxiv.2008.08694, sym_vqe_pg}, also resulting in lower two-qubit gate count.

One particular example where quantum computing can help is the simulation of refrigerants. 
The design of novel refrigerants has proven challenging due to trade-offs in key molecular properties 
such as global warming potential and ozone depletion potential, whilst also considering other properties 
such as toxicity, flammability, and stability. As such, molecular simulations have become increasingly used 
for investigating the structure-activity relationship of candidate refrigerants 
\cite{doi:10.1073/pnas.0913590107}. As a test case, we consider methane (CH$_4$) and its reaction with the 
hydroxyl radical (OH$^{\cdot}$),  
\begin{equation}
    \label{eq::atmospheric_reaction}
    \textrm{CH}_4 + \textrm{OH}^{\cdot} \rightarrow 
    [\textrm{CH}_3^{\cdot} \textrm{---} \textrm{H} \textrm{---} \textrm{OH}]^{\ddagger} \rightarrow 
    \textrm{CH}_3^{\cdot} + \textrm{H}_2\textrm{O}.
\end{equation}
\noindent
CH$_4$'s atmospheric properties have been thoroughly studied \cite{WUEBBLES2002177} and can provide guidelines 
in the application of quantum computers for search of new environment-friendly refrigerants. In order to 
characterise this process, one also needs to calculate a series of energies corresponding to the products, methyl 
radical (CH$_3^{\cdot}$) and water (H$_2$O). To estimate the reaction barrier that governs the kinetics of the 
reaction, we also simulate the transition state, 
$[\textrm{CH}_3^{\cdot} \textrm{---} \textrm{H} \textrm{---} \textrm{OH}]^{\ddagger}$.

In Section II, we give a detailed overview combining excitation filtering based on $Z_2$ symmetries, 
hard-core Boson representation, and favourable two-qubit gate cancellation (via Pauli-gadget synthesis scheme of 
Ref. \cite{cowtan2020generic}). Section III is devoted to the results obtained with our state-preparation strategy 
on System Model H1, powered by Honeywell. We focus on the resource reduction, and consider noisy calculations for 
CH$_4$'s optical spectra. Finally, we complete our investigation with a simulation of 
$[\textrm{CH}_3^{\cdot} \textrm{---} \textrm{H} \textrm{---} \textrm{OH}]^{\ddagger}$ and other reaction participants. 
We also apply symmetry verification to our calculations \cite{https://doi.org/10.48550/arxiv.2109.08401}. The two symmetry 
verification techniques we use are Partition Measurement Symmetry Verification (PMSV) and Mid-circuit Measurement Symmetry 
Verification (MMSV) \cite{Bonet_Monroig_2018}.

\begin{figure*}[!ht]
    \centering
    \includegraphics[width=0.95\linewidth]{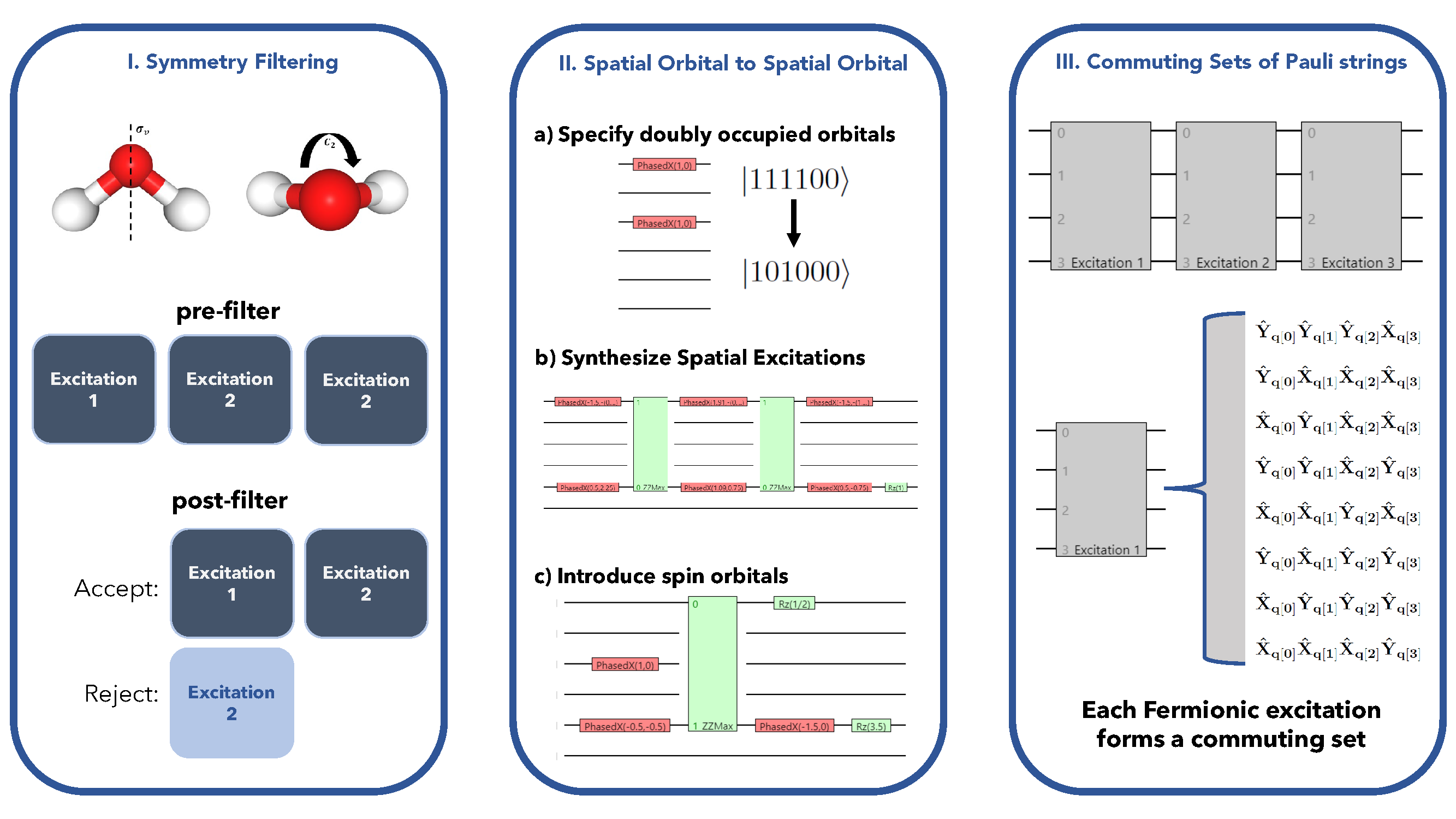}
    \caption{
        Schematic showing three major steps of the chemically aware ucc 
        state-preparation strategy. Step (i) uses symmetry as a priori to 
        discard excitations. Step (ii) is a compact synthesis scheme for 
        spatial to spatial excitations. Step (iii) uses tket to synthesise
        generic double and single excitations by commuting sets to maximize
        pauli-string cancellation. Each single or double UCC excitation is 
        naturally a commuting set of Pauli-strings.
    }
    \label{fig::efficient_uccsd_synth}
\end{figure*}

\section{Methods}
\subsection{Chemically Aware Unitary Coupled Cluster}
\label{sec::ca_ucc}
For our state-preparation strategy, we use Jordan-Wigner Encoding (JWE) to map Trotterised Fermionic exponents to 
Pauli operators acting on qubits \cite{wigner1928paulische, trotterization}. We refer to the by-product sequence of 
Pauli-$Z$ operations as JWE-strings, which consequently increase the effective $k$-locality of UCC exponents. Our 
spin orbitals and therefore the qubit register have alpha-beta ordering (each even-odd indexed spin orbital 
corresponds to a spatial orbital). Appendix \ref{appendix::synthesis} has more information on the UCC 
state-preparation method and conventional circuit decompositions (Individual and Commuting Sets synthesis). The steps 
of the chemically aware strategy are as follows:

\noindent
\textbf{I. Symmetry Filtering}: Filter the set of excitations composing the ansatz via use of molecular symmetry to identify forbidden terms. 
We used two techniques: 
\begin{enumerate}[(a)]
    \item Defining $Z_2$ symmetries to check commutation against the UCC 
    excitation operators \cite{Yen_2019, setia2020reducing, https://doi.org/10.48550/arxiv.2008.08694};
    \item A point group symmetry filtering method for CCSD adapted to be used for UCCSD \cite{carsky}.
\end{enumerate}

\noindent
\textbf{II. Spatial Orbital to Spatial Orbital}: There are three steps to compactly describe a pair of electrons excited between 
spatial orbitals. For this method, we necessarily change the ordering of the excitations to benefit from the two-qubit gate 
savings associated with spatial to spatial UCC excitation.
\begin{enumerate}[(a)]
    \item \textbf{Specify Double Occupied Spatial Orbitals}: Only doubly occupied ($\ket{1}$) and virtual spatial orbitals ($\ket{0}$) are 
    considered. For example, the Hartree-Fock state $\ket{111000}$ defining molecular spin orbital occupation, would be processed to return 
    $\ket{100}$ in the molecular spatial orbital occupation. These occupations are mapped to the even-indexed qubits on the circuit, 
    $\ket{100000}$. Single occupied spatial orbitals are ignored.
    
    \item \textbf{Hard-core Boson Representation}: Operations that excite a pair electrons from and to the same spatial orbital 
    can be synthesised more efficiently. 
    These excitations are of the type:
    \begin{equation}
        \label{eq::mo2mo_spin}
        \hat{a}^{\dagger}_{2p} \hat{a}_{2q} \hat{a}^{\dagger}_{2p+1} \hat{a}_{2q+1} - h.c., \\
    \end{equation}
    where $q < p$, and both variables track the spatial orbital index. Applying JWE results in 8 unique Pauli exponents over 4-qubits.
    
    These excitations lack JWE-strings, signifying zero parity exchange as these adjacent electrons hop between different spatial orbitals. 
    It can be seen that these excitations act on spin-orbitals, but excite and de-excite electrons between spatial orbitals. These adjacent 
    electrons travel together, yet cannot occupy the same spatial orbital with another pair of electrons. Equation \ref{eq::mo2mo_spin} can 
    also be expressed as,
    \begin{equation}
        \label{eq::mo2mo_spatial}
        \hat{b}^{\dagger}_{p} \hat{b}_{q} - h.c., \\
    \end{equation}
    where $\hat{b}$ denotes a Hard-core Bosonic operation. Eq. \ref{eq::mo2mo_spatial} can be re-expressed by using the equivalence between 
    Hard-core Bosons and Pauli operations \cite{matsubara, batyev_braginskii},
    \begin{equation}
        \label{eq::mo2mo_qubit}
        \frac{1j}{2} \left\{ \hat{Y}_{q} \hat{X}_{p} - \hat{X}_{q} \hat{Y}_{p} \right\}.
    \end{equation}
    We relabel the indices of equation \ref{eq::mo2mo_qubit} from $p \rightarrow 2p$ and $q \rightarrow 2q$. Each of these spatial orbital 
    to spatial orbital excitations act on 2 qubits and require 2 two-qubit gates.

    \item \textbf{Introduce Spin Orbitals}: We apply two-qubit gates on relevant even-odd qubits. Each even-indexed qubit 
    corresponds to the alpha spin-orbital of a spatial orbital, and similarly each beta spin-orbital is represented by an 
    odd-indexed qubit  \ref{eq::mo2mo_spatial}. single occupation of a spatial orbital is included by initializing the 
    relevant alpha-index qubit to the $\ket{1}$ state.
\end{enumerate}

\noindent
\textbf{III. Commuting sets of Pauli strings}: The remaining double and single excitations are synthesised. Each double excitation contains 8 Pauli-sub terms and these 
terms naturally form a commuting set. We synthesise these excitations in commuting sets with tket, resulting in 14 two-qubit gates at best. 
Increasing the length of JWE-strings increases the number of two-qubit gates for the corresponding decomposition. Circuits for single fermionic 
excitations contain at minimum four two-qubit gates and act over 3 qubits. Both two-qubit gate count and number of qubits grow as the number of Pauli-$Z$ 
strings increase.

\section{Results}
\label{sec::results}
\begin{figure*}
    \centering
    \includegraphics[width=\textwidth]{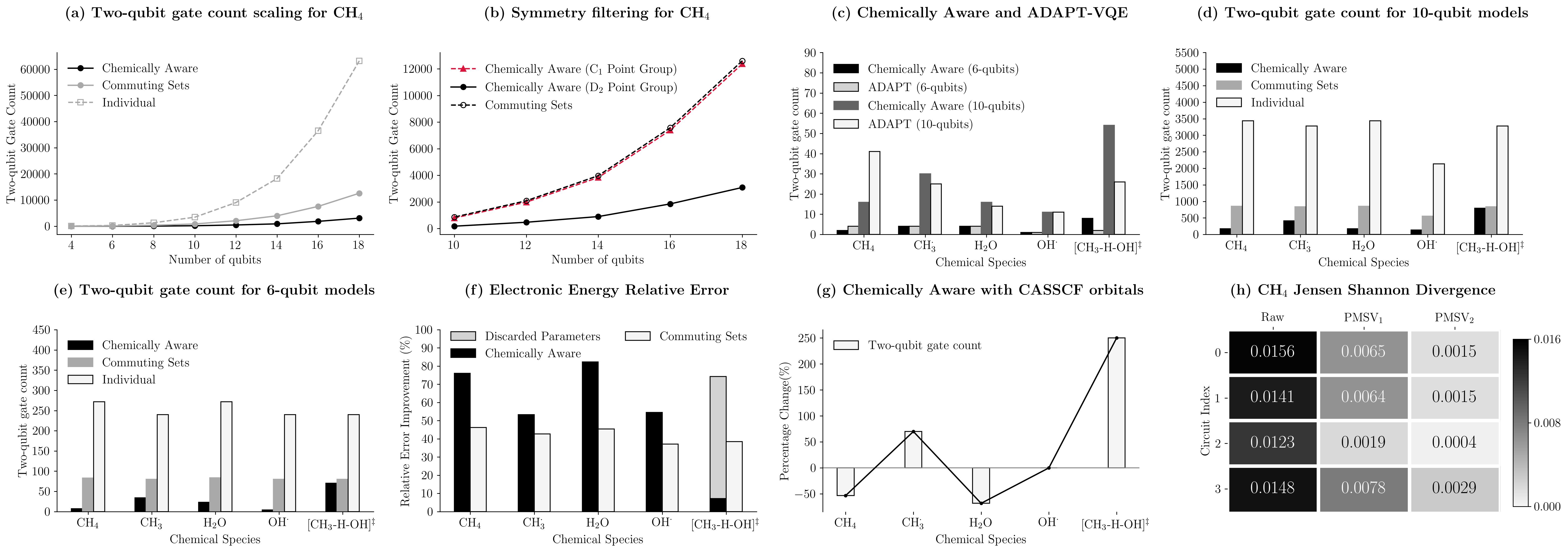}
    \caption{
        \textbf{(a)}: Two-qubit gate count scaling for various CH$_4$ active spaces. \textbf{(b)} Role of symmetry 
        in two-qubit gate count improvement coming from chemically aware. Test case CH$_4$ with $D_2$ point group. 
        \textbf{(c)} Comparison of excitation filtering in ADAPT compared to chemically aware for 10-qubit and 
        6-qubit models. \textbf{(d)} Two-qubit gate count comparison between chemically aware strategy, commuting sets 
        and individual synthesis circuits with symbolic gates for 10-qubits and \textbf{(e)} 6-qubits. \textbf{(f)} 
        Improvement in relative error on ground-state energy on H1-1E for chemically aware (commuting sets) compared to 
        individual synthesis. \textbf{(g)} Two-qubit gate count improvement using chemically aware over commuting sets 
        for CASSCF optimised 6-qubit models. \textbf{(h)} Jensen-Shannon Divergence for measurement statistics per 
        simulated chemically aware CH$_4$ circuit (indexed between 0 and 3) on H1-2 for three experiments. 
        Raw is a sampling experiment with no error mitigation, PMSV$_1$ is use of total electron conservation to post-select 
        measurement results, and PMSV$_2$ is use of all $U_1$ (alpha- and beta-electron number conservation) and $Z_2$ 
        operators (two symmetric transformations coming from point group of MOs).
    }
    \label{fig::ca_ucc}
\end{figure*}

All the circuits in this paper are prepared using Quantinuum's quantum chemistry package InQuanto \cite{inquanto, quantinuum}. 
The integrals to obtain the relevant chemistry Hamiltonians were obtained using an InQuanto extension to the open-source 
chemistry software package, PySCF, known as InQuanto-PySCF \cite{pyscf}. InQuanto provides the tools provide the pre- \& 
post-processing  logic to perform quantum computation, and more importantly to map the results from quantum computation back 
to the Chemistry problem. The UCC state-preparation techniques used in this paper are available in InQuanto. NGLView is used 
to visualize chemical structures and molecular orbitals via an InQuanto interface. We use the open-source ADCC library to 
compute benchmark optical spectra data at ADC-2 level of theory, \cite{https://doi.org/10.1002/wcms.1462}.

For this investigation, we used system model H1 devices and emulators \cite{Pino_2021, h1}. The IBMQ \textit{qasm} simulator was 
used to perform noiseless state-vector calculations, simulations with finite sampling noise and zero quantum noise. We use tket 
to synthesise, optimize and retarget quantum circuits to enable execution on a H1 hardware \cite{Sivarajah_2020, tket}. 

All our benchmark calculation are non-variational operator averaging calculations (Hamiltonian Averaging or QSE). The 
ground-state parameters for our state-preparation circuits are obtained via a noiseless VQE qasm simulation.

\subsection{Improvements due to the Chemically Aware Strategy}
We investigate the two-qubit gate count across three different methods, chemically aware, commuting sets, 
and individual UCCSD synthesis. We use CH$_4$ at equilibrium geometry as a benchmark system, alongside the 
$D_2$ point group to describe the molecular orbital symmetry \cite{NIST}. Fig. \ref{fig::ca_ucc}a reports 
an improvement by approximately 81\% (commuting sets) and 95\% (individual) in two-qubit gate count for various 
active-spaces ranging from 4 qubits to 18 qubits. By neglecting symmetry ($C_1$ point group), 
we reduce the efficacy of chemically aware to reduce two-qubit gate count, rendering only a small improvement 
due to the compact spatial-orbital-only excitation synthesis. Fig. \ref{fig::ca_ucc}b 
does not report any improvement in overall scaling with chemically aware compared 
to commuting sets synthesis for CH$_4$ with $C_1$ symmetry. 

ADAPT-VQE improves upon individual synthesis by iteratively constructing a compact 
ansatz. At the cost of increasing the total number of measurements, ADAPT discards both symmetry-forbidden and 
minimally contributing symmetry-allowed excitations. As a consequence, ADAPT 
circuits have a lower two-qubit gate count. With the chemically aware strategy, we bypass ADAPT's extra measurement cost 
to improve two-qubit gate count. Fig. \ref{fig::ca_ucc}c shows the number of Fermionic excitations with chemically aware 
compared to ADAPT. For symmetric molecules, ADAPT approximately returns a similar number of Fermionic excitations as 
chemically aware for both 6-qubit and 10-qubit active spaces. For 
$[\textrm{CH}_3^{\cdot} \textrm{---} H \textrm{---} \textrm{OH}^{\cdot}]^{\ddagger}$,  a 75\% (6-qubits) and 50\% (10-qubits) improvement 
in the number of excitations is observed. We attribute it to: (i) Additional symmetry in 
$[\textrm{CH}_3^{\cdot} \textrm{---} H \textrm{---} \textrm{OH}^{\cdot}]^{\ddagger}$ not exploited by chemically aware; 
(ii) ADAPT discards symmetry-allowed excitations of minimal contribution to the "true" ground state. 

For 10-qubit and 6-qubit models from Eq. \ref{eq::atmospheric_reaction}, we show empirical improvement in chemically aware two-qubit 
count compared to statistics from benchmark methods (commuting sets and individual). The two-qubit gate counts provided are for circuits 
with arbitrary phase gates with symbolic rotation parameters, hence independent from parameter sets characterizing specific physical states. 
Excluding the transition state and CH$_3^{\cdot}$, chemically aware improves upon commuting sets by 75\%-80\% (10-qubits) and 70\%-95\% (6-qubits). 
For less symmetric species, CH$_3^{\cdot}$ and $[\textrm{CH}_3^{\cdot} \textrm{---} H \textrm{---} \textrm{OH}^{\cdot}]^{\ddagger}$, 
we observe a less drastic two-qubit gate improvement with chemically aware.

With chemically aware state-preparation, we expect to see lower error in electronic energy
compared to calculations using commuting sets synthesis. We compare improvement in relative error using the chemically aware
strategy (commuting sets synthesis) compared to individual synthesis at ground-state parameters on the H1 emulator (H1-1E).
Excluding the transition state, Fig. \ref{fig::ca_ucc}f reports substantial improvement. For
$[\textrm{CH}_3^{\cdot} \textrm{---} H \textrm{---} \textrm{OH}^{\cdot}]^{\ddagger}$, chemically aware offers less improvement (7\%) 
compared to commuting sets (38\%). We note that tket defines arbitrary-phase gates as redundant for rotations 
below the $1 \times 10^{-12}$ threshold. The parameter set for commuting sets synthesised circuits 
(27 two-qubit gates) contains angles of size ~10$^{-30}$. For the chemically aware circuit (60 two-qubit gates), the
smallest parameters are of magnitude ~10$^{-7}$. Discarding parameters smaller than $1\times10^{-3}$ has a negligible impact 
on the electronic ground state energy, and we observe an improvement (74\%) in the chemically aware calculation 
(16 two-qubit gate count). 

Since we are using active space models, we also need to relax the molecular orbital coefficients within the defined 
active space via a Complete Active Space SCF (CASSCF) routine. This aids in better recovery of electronic correlation 
with VQE methods. Fig. \ref{fig::ca_ucc}g reports the two-qubit gate count increase for chemically aware when using 
CASSCF orbitals over SCF orbitals at  ground-state parameters. We only see an increase in two-qubit gate depth for 
CH$_3^{\cdot}$ and the transition state, mainly due to increase in the number of non-negligible UCC parameters 
(greater than ~$10^{-3}$). This leads to less excitations being redundant and therefore more two-qubit gates. 
All species (excluding OH$^{\cdot}$) report varying degrees of improvement in electronic correlation 
recovery. 

We perform Hamiltonian averaging experiments for 6-qubit CH$_4$ with $D_2$ symmetry using (i) no error mitigation, (ii) PMSV with 
a $U_1$ symmetry constraint (total electron conservation) and (iii) with alpha- and beta-number conservation and Point 
group operations defined as $Z_2$ symmetries. We require 4 circuits to measure the electronic energy. 
Tab. \ref{tab::ca_ucc_energies} in Appendix \ref{app::ch4_energy} reports improvement in relative error as we add more constraints to PMSV. 
Fig. \ref{fig::ca_ucc}h shows improvement in Jensen-Shannon Divergence (JSD) by one order of magnitude 
after PMSV application. We simulate a CH$_4$ circuit with 500 shots to generate measurement statistics on H1-2. The distribution of outcomes 
is compared with output from a noiseless qasm simulator. A JSD with value zero means experimental measurement distribution exactly agrees with 
theory data. After PMSV$_2$ application (see Fig. \ref{fig::ca_ucc}g for definition), 
we observe a JSD value of magnitude $~10^{-3}$.

\label{sec::ca_ucc_anal}

\subsection{Excited States \& Optical Spectra}
In order to improve feasibility for hardware, we employed a set of nearly 
orthonormal custom expansion operators. We observe a drastic 
reduction of resources at the cost of finding only a subset of singlet 
states and zero triplet states. Using a near orthonormal 
set of expansion operators causes less error to accumulate in the overlap 
quantity, $S$, of the QSE equation, $Hc = eSc$. We use a circuit with 7 two-qubit gates 
and sample 31 measurement circuits 500 times each. Tab. 
\ref{tab::qse_6q_experiment_2} successfully reports 4 out of 5 excitation energies. Each excitation 
is defined from the singlet ground-state, $S_0$, to a singlet excited state, $S_n$.
We use total-electron number conservation as a 
constraint to post-select on the measurement results. We find the penalty 
of adding extra two-qubit gates renders MMSV ineffective, giving a similar level of accuracy as 
the experiment with no error mitigation (Raw). PMSV is effective at reducing ground state deviation. 
Similar to the Hamiltonian averaging procedure, 
using more symmetry constraints (PMSV$_2$) improves accuracy of the computed quantity. 
With PMSV$_2$, we are able to replicate degeneracy between $S_2$ and $S_3$ (to three decimal places). We 
note with PMSV application, the excitation energies are increasing overestimated from $S_1$ to $S_4$.

\begin{table}[!ht]
	\centering
	\caption{
		Excitation energies from ground state singlet, $S_0$, to the first four excited states, 
		$S_n$, where $n=\left\{0, 1, 2, 3, 4, 5\right\}$ 
		QSE calculation on CH$_4$ at equilibrium geometry on H1-2 hardware. MMSV, PMSV$_1$ 
		and PMSV$_2$ used to mitigate errors. EOM
		is EOM-CCSD with intel x86$\_$64 processor.
	}
	\small{
	\begin{tabular}{|c|c|c|c|c|c|}
		\hline
		{} & \small{\textbf{MMSV}} & \small{\textbf{PMSV}$\mathbf{_1}$} & \small{\textbf{PMSV}$\mathbf{_2}$} 
		& \small{\textbf{Raw}} & \small{\textbf{EOM}} \\
		{} & \small{(Ha)} & \small{(Ha)} & \small{(Ha)} & \small{(Ha)} & \small{(Ha)} \\
		\hline
		\small{$\mathbf{S_0 \rightarrow S_1}$} &  0.837 & 0.843 & 0.868 & 0.836 &  0.862 \\
		\hline
		\small{$\mathbf{S_0 \rightarrow S_2}$} &  0.854 & 0.852 & 0.868 & 0.848 &  0.862 \\
		\hline
		\small{$\mathbf{S_0 \rightarrow S_3}$} &  1.616 & 1.617 & 1.639 & 1.608 &  1.634 \\
		\hline
		\small{$\mathbf{S_0 \rightarrow S_4}$} &  1.718 & 1.735 & 1.749 & 1.724 &  1.715 \\
		\hline
		\small{$\mathbf{S_0 \rightarrow S_5}$} & - & - & - & - & 1.740 \\
		\hline
	\end{tabular}
	}
	\label{tab::qse_6q_experiment_2}
\end{table}

After performing the QSE computation, we obtain a description of a set of excited states, 
\begin{equation}
	\ket{\psi_v} = \sum_k w_k^v \hat{F}_k \ket{\psi_0},
	\label{eq::qse_vector}
\end{equation}
where $\ket{\psi_0}$ is the ground-state, $w_k^v$ is a coefficient vector for the $v^{th}$ excited state found with 
QSE and $\hat{F}_k$ are the corresponding QSE expansion operators (i.e. $\hat{a}^{\dagger}_2 \hat{a}_0$). With this 
information post-QSE procedure, one can cast an overlap, $\bra{\psi_0}\ket{\psi_v}$, into a sequence of operator averaging 
procedures, $\sum_k \bra{\psi_0} w_k^v \hat{F}_k \ket{\psi_0}$. This is extremely useful for evaluating quantities needed to 
obtain optical spectra, i.e. transition density matrices and transition dipole moments. 

Our intention is to obtain optical spectra with the QSE result. We use dipole operators, $\hat{\mu}_{\alpha}$, with $\alpha=(x, y, z)$ 
and the origin as the charge centre. A post-QSE operator averaging procedure is performed to obtain transition dipole moments for 
each of the singlet states, $\bra{\psi_0} \hat{\mu}_{\alpha} \ket{\psi_v}$, using Eq. \ref{eq::qse_vector} to absorb the 
QSE expansion operators into $\hat{\mu}_{\alpha}$. The oscillator strength, $f$, can be obtained as a post-processing 
procedure,
\begin{equation}
	f = \frac{2 \epsilon_v}{3} \sum_{\alpha=\left\{ x, y, z \right\}} \left| \bra{\psi_0} \hat{\mu}_{\alpha} \ket{\psi_v} \right|^2,
\end{equation}
where $\epsilon_v$ is the transition energy from state $\ket{\psi_0}$ to $\ket{\psi_v}$. 

\begin{figure}[!ht]
    \centering
	\includegraphics[width=0.45\textwidth]{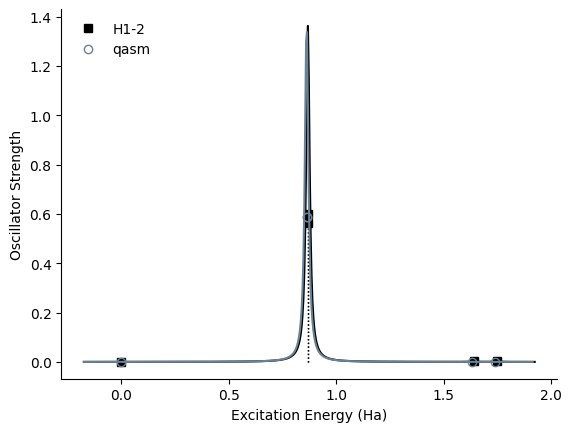}
    \caption{
		Optical spectra calculated with H1-2 hardware (black squares) 
		and noiseless qasm simulator (blue circles). 
		Black line is Lorentzian line broadening.
		}
    \label{fig::optical spectra}
\end{figure}

Fig. \ref{fig::optical spectra} reports optical spectra data computed with H1-2 hardware. A subsequent transition dipole moment 
calculation was performed on H1-2 with 500 shots per circuit. Comparing with noiseless qasm simulations as a benchmark, 
there is one peak at 0.86 $\si{\haunit}$ (oscillator strength 0.59). This peak corresponds to two degenerate states. 
The H1-2 quantum computer produces two peaks very close at 0.8676 $\si{\haunit}$ (oscillator strength 0.601) 
and 0.8684 $\si{\haunit}$ (oscillator strength 0.565). This small discrepancy between two degenerate states 
is added primarily by sampling noise. 

\label{sec::excited_states}

\subsection{Atmospheric Reaction Simulation}
After presenting our results for the characterization of the methane molecule, we proceed now to show our results 
for the simulation of the CH$_4$ + OH$^\cdot$ $\rightarrow$ CH$_3^{\cdot}$ + H$_2$O reaction. 
We focus first on the thermodynamics of reactants and products, and then proceed to analyse our 
results for the transition state and the activation energy for the reaction. Here, we present our results 
for 6-qubit active space models, which are suitable for experiments on hardware. 

The results for the activation energy of the reaction are shown in Table \ref{tab:barrierss}. 
The absolute value of the energy differences is too small to be reproduced, which can be seen 
from the results of the noiseless simulation. The main reason for this is the small number of 
qubits considered and, as a consequence, insufficient accuracy for each of the species 
participating in the reaction. As a result, in a hardware experiment the activation energy is 
found to be negative, although the order of magnitude of its absolute value is correct for the 
latter. This result is an indicator of the general difficulty of accurately calculating reaction 
barriers, due to the small energy differences involved and large accuracy required. To our 
knowledge, this is the first time an activation energy has been calculated on an actual ion 
trap hardware although theoretical work estimating reaction energy profiles using much larger 
numbers of qubits exist \cite{Reiher_2017}.

\begin{table}[H]
    \centering
    \caption{
        Activation energy for the reaction between methane and the hydroxyl radical ({AE}), 
        and for the reverse process ({R-AE}).
    }
    \label{tab:barrierss}
    \small{\begin{tabular}{|l|c|c|c|}
    \hline
    {} &  \textbf{Ref. \cite{espinosagarcia_2015_qct}} & \textbf{qasm} &  \textbf{H1-1} \\
    {} & \textbf{(\si{\haunit})} & \textbf{(\si{\haunit})} & \textbf{(\si{\haunit})} \\
    \hline
    AE &  +0.01  &  +0.12  &  -0.05 \\
    \hline
    R-AE  &  +0.02  & +0.12    &  +0.01 \\
    \hline
    \end{tabular}}
\end{table}

\label{sec::atm_react_sim}

\section{Discussion}
In this work, we have: (a) improved UCC circuit synthesis to reduce two-qubit gate count, (b) 
performed a 6-qubit QSE calculation on System Model H1 to output optical spectra (c) estimated the 
reaction barrier from  Eq. \ref{eq::atmospheric_reaction}.

Alongside the two-qubit gate savings from state-preparation, the PMSV error mitigation 
technique has allowed us to significantly improve the quality of the energy calculations 
by reducing absolute error on the ground-state energy. For CH$_4$ and OH$^{\cdot}$, 
the results are in strong agreement with state-vector simulations. Alongside the 6-qubit 
QSE calculation showcased by Ref. \cite{https://doi.org/10.48550/arxiv.2208.02414} on
 a superconducting IBMQ quantum computer, we have performed a 6-qubit excited states calculation 
for methane on ion-trap hardware (System Model H1), advancing the industry-standard of using 
``4 qubits or less'' \cite{schrodinger_table}. We note that symmetry-verification is 
applicable to small molecules with number conservation properties and mirror-planes, 
and should be useful for relevant future experiments. However, we note the efficacy 
of PMSV may deteriorate as we scale to larger, more non-symmetric molecules that require 
a greater number of qubits. The long term need for error correction still exists.

We also employed H1's mid-circuit measurement facility to symmetry verify our calculation results. 
Whilst outperformed by PMSV, it is still inconclusive to exclude the use of 
mid-circuit measurements. We believe this functionality still has more applications 
for hardware chemistry experiments, i.e. approximating required quantities in variational 
real time evolution.

Computation of chemical quantities, such as UV/ VS spectra is essential for comparison with 
experimental data. We use the QSE description of the excited states to evaluate 
transition dipole moments and ultimately optical spectra. We learn optical spectra quality 
depends on excitation energy accuracy of the prior QSE procedure. Most importantly, we control 
sampling error from the prior QSE procedure, by choosing select single and double UCC 
excitations, albeit at the cost of restricting to a subset of singlet excited states.

From our investigation to compute the reaction barrier, we learn that dynamical correlation is difficult 
to recover on today's quantum computers. This is especially important when the energy difference between 
the reactants, products, and transition states are expected to be small. Specifically, to recover 
correlation for $[\textrm{CH}_3 \textrm{---} \textrm{H} \textrm{---} \textrm{OH}]^{\ddagger}$, 
many orbitals are needed. With chemically aware UCC Synthesis, the decrease in two-qubit gate count was 
not as dramatic as the other species, due to molecular asymmetry. A slight improvement in two-qubit gate 
count is observed and is outperformed by Fermionic ADAPT VQE. We believe that there maybe additional 
unexploited redundancy in our model of $[\textrm{CH}_3 \textrm{---} \textrm{H} \textrm{---} \textrm{OH}]^{\ddagger}$. 
However, the possibility remains that ADAPT discards symmetry-allowed unimportant excitations. Further 
analysis is needed to investigate this and to determine if the filtering by ADAPT can be performed as a 
priori on a classical computer. We also did not investigate the use of ADAPT-VQE as a second UCC 
excitation filter in the chemically aware strategy. We dedicate future work towards this effort.

\section{Conclusion}
We incrementally extend the reach of quantum computation by performing the QSE algorithm to obtain 
chemically relevant quantities such as optical spectra. Open questions remain in the use of QSE to 
obtain vibrational or infrared spectra.

Alongside two-qubit gate count, the number of measurements is a major bottleneck 
preventing scaling beyond 10-qubits. We have identified the need to drastically reduce the cost of 
the operator averaging procedure, which is used in VQE and QSE. Follow-up analysis of newer measurement 
methods is necessary \cite{Zhao_2021, https://doi.org/10.48550/arxiv.2208.14490}. We additionally 
believe that run-time requirements and accuracy for QSE can be reduced by application of point-group 
symmetry. This is done similarly for EOM-CCSD, and the application to QSE needs to be investigated in 
detail. 

To summarize, we have showed in this paper a series of techniques to help extract the maximum yield 
from current and near-term hardware. We anticipate that these techniques will help enable future small 
chemistry experiments on quantum devices on near-term hardware.

\footnotesize{
    \subsubsection*{Acknowledgements}
    We would like to thank Quantinuum's hardware \& software units for discussions 
surrounding excited state chemistry, experiment design, results analysis and device 
access. Specifically, we thank Michal Krompiec, David Zsolt Manrique, Kentaro Yamamoto,  
Jenni Strabley, Brian Neyenhuis, Simon McAdams, Ross Duncan, Seyon Sivarajah, 
John Children, Silas Dilkes, Alec Edgington, Sam White, Isobel Hooper,
Andrew Tranter, Moshin Iqbal, Henrik Dreyer \& Gavin 
Towler. We are also grateful to Joaquin Espinosa-Garcia and Jose C. Corchado for 
providing the geometrical configuration for 
$[\textrm{CH}_3 \textrm{---} \textrm{H} \textrm{---} \textrm{OH}]^{\ddagger}$. 
We appreciate positively the support provided by Honeywell 
Performance Materials \& Technology (PMT).

    \subsubsection*{Author Contributions}
    \textbf{I. T. Khan} developed and investigated the chemically aware circuit state-preparation, 
and its application to refrigerant molecules. I. T. K also evaluated optical spectra via Quantum Subspace 
Expansion on System Model H1. \textbf{M. Tudorovskaya} peformed the reaction barrier study with 
calculations on System Model H1. \textbf{J. J. M. Kirsopp} implemented point group filtering 
from Ref. \cite{carsky}. \textbf{D. Mu\~{n}oz Ramo} provided technical oversight on the relevance 
and application of the methods outlined in this paper to the problem of atmospheric refrigerants. 
\textbf{P. Warrier, D. K. Papanastasiou \& R. Singh} defined the atmospheric refrigerants 
use-case, and provided oversight on application and relevance to quantum computing.

    \subsubsection*{Availability of Data and Materials}
    Data is available upon request from authors. All inquiries regarding the InQuanto Platform should be made to \href{mailto:inquanto-support@quantinuum.com}{inquanto-support@quantinuum.com}.

    \subsubsection*{Correspondence}
    Queries and requests should be adressed to Irfan T. Khan (\href{mailto:irfan.khan@quantinuum.com}{irfan.khan@quantinuum.com}), or David Mu\~{n}oz Ramo (\href{mailto:david.munozramo@quantinuum.com}{david.munozramo@quantinuum.com}). 

    \subsubsection*{Competing Interests}
    We declare the use of Quantinuum H-Series devices \& the IBMQ \textit{qasm} simulator. The views expressed in this paper do not reflect the official policy or position of IBM, Honeywell International and any related subsidiaries. IBM and Honeywell International are also Quantinuum shareholders.
}

\bibliography{sample}
\clearpage
\pagebreak
\newpage
\appendix
\vfill
\clearpage
\pagebreak
\newpage
\section{UCC Synthesis Methods}
\subsection*{UCCSD State-Preparation Methods}
The wavefunction describing each molecular system is prepared via a parameterised quantum circuit or ansatz that 
provides an appropriate description of electron correlation. We use the UCCSD operator. It is represented as a 
Trotterized product of exponents \cite{trotterization}, and mapped to Pauli representation using Jordan$-$Wigner 
encoding (JWE) \cite{wigner1928paulische, Aspuru_Guzik_2005}. With JWE, each qubit represents the electron number 
occupation of each spin-orbital in our system - $\ket{1}$ is occupied and $\ket{0}$ is unoccupied. We use 
alpha-beta ($ab$) ordering, so that spatial orbital index $p$ corresponds to even-indexed and odd-indexed 
spin-orbitals, $2p$ and $2p+1$. Using JWE leads to a sequence of Pauli-$Z$ operations, 
necessary to preserve antisymmetry of our UCC state, but at the cost of increasing effective $k$-locality. 
We refer to these Pauli-$Z$ operations as JWE-strings.

The UCCSD operator is Trotterized and written as a product of Pauli
exponentials via JWE:
\begin{equation}
e^{\hat{T} - \hat{T}^{\dagger}} \approx \prod_{m} \prod_{n} e^{i \theta_{m} \hat{P}_{m,n}},
\label{trotterization}
\end{equation}
where the index $m$ runs over each distinct Fermionic excitation, and index $n$ runs over each Pauli-sub-term, $\hat{P}_{m,n}$, of 
that Fermionic excitation. In Eq. \ref{trotterization}, ${\theta}_{m}$ are independent real parameters, and $\hat{T}^{\dagger}$ is 
the sum over each excitation operator acting across all orbitals. The excitations that are included in the ansatz are determined by 
the spin-multiplicity, number of electrons and the basis-set size of the chemical system. UCCSD can be very costly in the number of 
two-qubit gates which has limited its practicality in hardware experiments.

The chemically aware UCC synthesis is described in the methods section. A sumary is provided below: 
\begin{itemize}
    \item A symmetry filtering step to discard redundant Fermionic excitations \cite{Yen_2019}.
    \item A compact synthesis of Fermionic spatial-to-spatial excitations by treating them as Bosonic operations and applying 
    Jordan-Wigner encoding \cite{nam2019groundstate}. For each spatial-to-spatial excitation, the two-qubit gate count is reduced from 64 to 2 
    \cite{Cowtan_2020}.
    \item A Pauli-Gadget synthesis method leveraging commuting property of observables, available in tket \cite{cowtan2020generic}. 
    This is used to synthesise generic double UCC excitations and single UCC excitations.
\end{itemize}

We compare the chemically aware state-preparation with commuting sets and individual synthesis in 
Sec. \ref{sec::ca_ucc_anal}. A description for the latter two methods is provided below.

\subsection*{Commuting Sets}
A UCC synthesis method leveraging commuting property of observables, available in tket, \cite{cowtan2020generic}. 
The full set of single and double UCC excitations are generated for a specific Fermionic reference state, no 
additional filtering based on symmetry or any other chemical analysis is performed. The default ordering of the 
UCC excitations is not changed, i.e. singles are synthesised before doubles. Within our study, we use each UCC 
excitation as its own distinct commuting set. UCC double excitations with length zero JWE-strings can be synthesised 
with 14 two-qubit gates. Excitations with arbitrary length JWE-strings have a two-qubit count greater than 14.

The Pauli-sub-terms within the ansatz can be intermixed to achieve more favourable two-qubit gate cancellation 
(caused by JW-strings), but this is not performed here. We are not sure how this would affect Trotter error.

\subsection*{Individual}
\begin{figure}[H]
    \includegraphics[width=\linewidth]{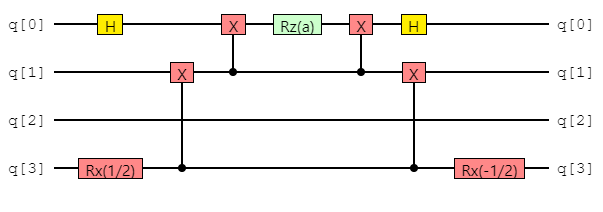}
    \caption{A circuit primitive for $e^{-\pi/2 \theta \hat{P}}$, where $\hat{P}$ is $\hat{Y}_0 \hat{X}_1 \hat{Z}_2$. Circuit is 
    visualised with tket. }
    \label{fig::naive_ucc_primitive}
\end{figure}

Brute-force synthesis of UCC excitations. Each Pauli-sub-term for each Fermionic excitation is synthesised with the Hadamard 
gate, $\hat{R}_x$ with fixed gate rotation $\pm \pi/4$, a cascade of two-qubit gates and a arbitrary phase $\hat{R_z} (\theta)$ 
rotation gate. A circuit primitive for $e^{-\pi/2 \theta \hat{Y}_0 \hat{X}_1 \hat{Z}_2}$ is shown in Fig. \ref{fig::naive_ucc_primitive}. 
Ref. \cite{Whitfield_2011} and Ref. \cite{romero2018strategies} provide more details on this.
\label{appendix::synthesis}

\section{CH$_4$ 6-qubit active space model}
\begin{figure}[H]
    \includegraphics[width=\linewidth]{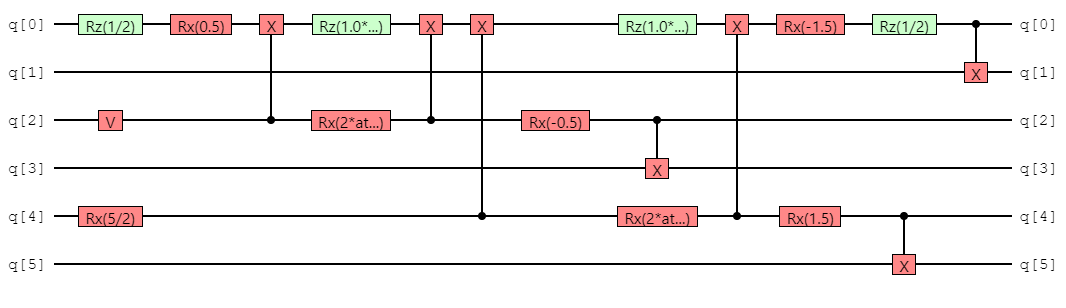}
    \caption{
        Chemically aware UCC circuit containing only two spatial to spatial orbital double excitations. 
        The other double and single excitations are discarded as part of the chemically aware procedure. 
    }
    \label{fig::ucc_ch4}
\end{figure}

InQuanto-PySCF describes MOs of CH$_4$ with the point group $D_2$. In the active-space, the occupied orbital has 
irrep $b_3$, and two degenerate virtual orbitals have irrep $b_1$ and $b_2$. There are two $U_1$ symmetry 
operators corresponding to alpha- and beta-electron number conservation. There are two further operations 
corresponding to $\pi$ radian rotations of the CH$_4$ molecule around $z$- and $y$-axes:
\begin{itemize}
    \item $\hat{C}_z (\pi/2): \hat{Z}_0 \hat{Z}_1 \hat{Z}_2 \hat{Z}_3$;
    \item $\hat{C}_y (\pi/2): \hat{Z}_0 \hat{Z}_1 \hat{Z}_4 \hat{Z}_5$.
\end{itemize}
We employ these symmetries in the chemically aware strategy resulting in only two spatial to spatial doubles 
(two-qubit gate count 7). The circuit used for CH$_4$ calculations is shown in Fig. \ref{fig::ucc_ch4}.

For the calculations in this paper, we built chemical models with InQuanto-PySCF for the following reaction 
participants, CH$_4$, CH$_3^{\cdot}$, H$_2$O, OH$^{\cdot}$, and [CH$_3^{\cdot}$ $\textendash$ H $\textendash$ OH]$^{\ddagger}$. 
We used minimal basis and neutral electrostatic charge for all species. For the closed-shell (open-shell) species we 
used an active space of three spatial orbitals and two (three) electrons. CASSCF was used to optimize for the active 
space of choice. For certain species, the guess point group assigned by PySCF affected the amount of correlation 
recoverable with UCC-VQE or Post-HF methods (CCSD). In Tab. \ref{tab::model_correlation_confession}, we report 
electronic correlation energy for CH$_4$ when using $C_1$ or $D_2$ point groups with (CASSCF orbitals) and 
without (SCF orbitals) orbital optimization.

\begin{table}[H]
    \caption{
        Table showing correlation energy (Ha) in CH$_4$ when using CASSCF 
        and SCF orbitals with D$_2$ or C$_1$ point group symmetry.
    }
    \label{tab::model_correlation_confession}
    \begin{tabular}{|c|c|c|}
        \hline
        Point Group & CASSCF Orbitals & SCF Orbitals \\
        \hline
        $D_2$ & -0.0026 & -0.0026 \\
        \hline
        $C_1$ & -0.0158 & -0.0105 \\
        \hline
    \end{tabular}
\end{table}

\label{appendix::circuits}

\section{Trotter Error in CH$_3$ 6-qubit Chemically Aware UCC Circuit}
For CH$_3$ after parametric compilation, we note an initial decrease from 88 two-qubit gates to 50 two-qubit gates for 
the chemically aware circuit. This is due to the excitation reordering step which also changes the optimal 
parameters for those excitations. To achieve the two-qubit gate count of 20 for CH$_3$ with chemically aware, we defined 
an absolute tolerance ($1 \times 10^{-4}$) to discard small excitations 
($\hat{a}_4^{\dagger} \hat{a}_2 \hat{a}_3^{\dagger} \hat{a}_1 - h.c.$). We report the subset of excitations synthesised 
on the chemically aware and commuting sets circuit in Tab. \ref{tab::ch3_excitations}.

\begin{table}[H]
    \caption{
        Showing CH$_3$ UCC excitations with their ordering and optimal parameters on 
        chemically aware$^{\dagger}$ and commuting sets$^{\bot}$ circuit. Optimal 
        parameters defined as UCC angles in radians.
    }
    \label{tab::ch3_excitations}
    \begin{tabular}{|l|c|c|c|c|}
        \hline
        \textbf{Exponents} & \textbf{Order$^{\dagger}$} & \textbf{Order$^{\bot}$} & \textbf{Angle$^{\dagger}$} & \textbf{Angle$^{\bot}$} \\
        \hline
        $\hat{a}_4^{\dagger} \hat{a}_0 \hat{a}_5^{\dagger} \hat{a}_1 - h.c.$ & 1 & 6 & -0.03611 & -0.07243 \\
        \hline
        $\hat{a}_4^{\dagger} \hat{a}_2 \hat{a}_3^{\dagger} \hat{a}_1 - h.c.$ & 2 & 7 & -0.00009 & 0 \\
        \hline
        $\hat{a}_4^{\dagger} \hat{a}_0 - h.c.$ & 3 & 0 & 0.00117 & 0 \\
        \hline
        $\hat{a}_5^{\dagger} \hat{a}_1 - h.c.$ & 4 & 1 & 0.00132 & 0 \\
        \hline
    \end{tabular}
\end{table}
\label{appendix::ch3_trotter_error}

\section{Symmetry Verification}
\label{verif_appendix}
\subsection*{Partition Measurement Symmetry Verification}
\begin{figure}[H]
    \centering
    \includegraphics[width=\linewidth]{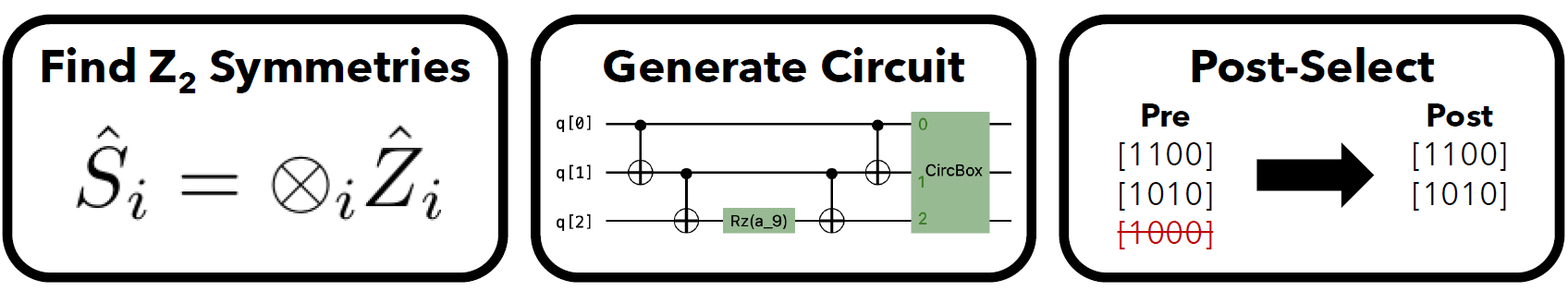}
    \caption{
        Schematic detailing pre- and post-processing steps to execute Partition Measurement Symmetry Verification.
    }
\end{figure}

The aim of PMSV (Partition Measurement Symmetry Verification) is to apply a constraint to the measurement result of some quantum 
computation that can be used in classical post-processing to improve upon the quality of calculation. Certain physical symmetries 
can be represented as Pauli-symmetries by keeping track of parity. In this paper, we exploit $Z_2$ symmetries such as mirror planes 
or unit cell translations and $U_1$ symmetries such as electron number conservation 
\cite{setia2020reducing, Yen_2019, https://doi.org/10.48550/arxiv.2008.08694}. If applicable to the problem, PMSV can be performed 
without increasing the number of measurement circuits. There is a small resource increase to measure the additional Pauli-symmetries, 
either extra two-qubit gates as part of the diagonalization process in measurement reduction, or measuring more qubits for a 
particular measurement circuit. We state the recipe for PMSV here:
\begin{enumerate}[(1)]
    \item Find largest Abelian point group of molecule. Transformations of this point group are Pauli-symmetries.
    \item Build symmetry verifiable circuits using the measurement reduction facility in tket. The operators that we need to 
    measure on the quantum device to solve out problem consist of Pauli-Is, Pauli-Xs, Pauli-Ys and Pauli-Zs across the qubit register. 
    These operations can be partitioned into commuting sets. For element in a commuting set, if the Pauli-symmetries commute 
    element-wise, it can be added to the commuting set. This way, each commuting set is symmetry verifiable. 
    \item Post-select on measurement result before counting bit-strings to compute expectation value of problem Hamiltonian. 
    The quantity to post-select on is the XOR sum over the bit-strings needed to compute expectation value of Pauli-symmetry.
\end{enumerate}

\subsection*{Mid-circuit Measurement Symmetry Verification}
\begin{figure}[H]
    \centering
    \includegraphics[width=\linewidth]{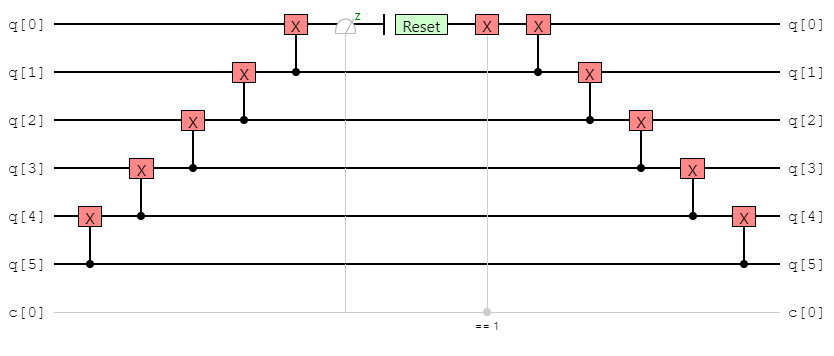}
    \caption{
        Circuit primitive to project into the -1 sector of the Pauli Symmetry $\hat{Z}_0 \hat{Z}_1 \hat{Z}_2 \hat{Z}_3 \hat{Z}_4 \hat{Z}_5$.
    }
    \label{fig::mcmsv_ciruit0}
\end{figure}
Mid-measurement symmetry-verification checks if a symmetry has been violated during runtime of the quantum circuit, 
and discards measurement results based on this check \cite{Bonet_Monroig_2018}. This technique has a 2-qubit gate penalty. 
For this problem, mid-circuit measurements were used to enforce total electron number conservation: adding 10 two-qubit gates 
to the measurement circuits.

Measuring the qubit in the middle leaves the qubit in a non-computational basis. Hence, we reset the qubit to the state, 
$\ket{0}$, followed by a classically conditioned reset on the classical bit, $c[0]$, to allow further simulation of the 
quantum circuit. To perform the projection, 
\begin{equation}
    \hat{P} = \hat{I} + \left( -1 \right) ^{x} \hat{S}
\end{equation}
normally one requires measuring each qubit. $\hat{S}$ is the Pauli-Symmetry and $x$ is the parity of the bitstring to accept or reject. 
We can transfer this measurement to only one qubit by using a CX gate cascade. The circuit primitive we use to project the state into a 
particular symmetry sector is given in Fig. \ref{fig::mcmsv_ciruit0}. We note that to symmetry verify each measurement circuit, 
the conflicting sets of Pauli-strings need to be synthesized into measurement circuit, hence introducing additional two-qubit gate. 
For the systems studied, the additional number of two-qubit gates is very small. 

\label{mid_measure}

\section{Data from Experiments}
\subsection*{Electronic ground state energy of CH$_4$ computed H1-2}
\label{app::ch4_energy}
Data from the H1-2 ground-state calculation in Sec. \ref{sec::ca_ucc_anal}. Jensen-Shannon 
Divergence of the measured distributions corresponding to the simulated circuits are also 
displayed in Fig. \ref{fig::ca_ucc}

\begin{table}[H]
    \centering
    \caption{
        Electronic Energy of CH$_4$ with 
        6-qubit active space and chemically 
        aware state-preparation circuit computed on H1-2.
    }
    \label{tab::ca_ucc_energies}
    \begin{tabular}{|c|c|c|}
        \hline
        {} & \textbf{Energy (Ha)} & \textbf{Rel. Error (\%)} \\
        \hline
        \textbf{Raw} & -39.656 & 0.184 \\
        \hline
        \textbf{PMSV}$\mathbf{_1}$ & -39.695 & 0.086 \\
        \hline
        \textbf{PMSV}$\mathbf{_2}$ & -39.721 & 0.020 \\
        \hline
    \end{tabular}
\end{table}

\subsection*{Ground state energies of the refrigerant molecules}
\label{app::refrigerant_energies}
Data behind the reaction barrier calculation in Sec. \ref{sec::atm_react_sim}, Tab. \ref{tab:barrierss}. 
\begin{table}[H]
    \centering
    \caption{Results from Hamiltonian averaging experiments on H1 and H1E. All molecules 
    are simulated on H1-1, except for CH$_4$ which is simulated on H1-2.
    Parameters characterizing ground-state are found via a noiseless VQE execution on \textit{qasm} simulator. 
    Noiseless energy from \textit{qasm} is also displayed as a benchmark.}
    \label{tab:exp_results}
    \begin{tabular}{|c|c|c|c|}
    \hline
    \textbf{Molecule} & \textbf{H1-1E} & \textbf{H1} & \textbf{qasm} \\
    {} & \textbf{(\si{\haunit})} & \textbf{(\si{\haunit})} & \textbf{(\si{\haunit})} \\
    \hline
    OH & -74.34 & -74.29 & -74.36  \\
    \hline
    \small{H$_2$O} & -74.92 & -74.77 & -74.97   \\
    \hline
    \small{CH$_3$} & -39.05 & -38.94 & -39.08  \\
    \hline
    \small{CH$_4$} & -39.72 & -39.72 & -39.73 \\
    \hline
    $[\textrm{CH}_3 \textrm{---} \textrm{H} \textrm{---} \textrm{OH}]^{\ddagger}$ & -113.98 & -113.77 & -114.04 \\
    \hline
    \end{tabular}
\end{table}

\end{document}